\documentclass[manuscript,screen]{acmart}
\AtBeginDocument{%
  }

\setcopyright{acmlicensed}
\copyrightyear{2025}
\acmYear{2025}
\acmDOI{XXXXXXX.XXXXXXX}
\acmConference[CUI '25]{ACM Conference on Conversational User Interfaces Workshop on Personas Evolved: Designing Ethical LLM-Based Conversational Agent Personalities}{July 8-10}{Waterloo, Canada}

\begin{document}

\title{Promoting Online Safety by Simulating Unsafe Conversations with LLMs}

\author{Owen Hoffman}
\email{}
\affiliation{%
  \institution{Department of Computer Science, Swarthmore College}
  \city{Swarthmore}
  \state{Pennsylvania}
  \country{USA}
}

\author{Kangze Peng}
\email{}
\affiliation{%
  \institution{Department of Computer Science, Swarthmore College}
  \city{Swarthmore}
  \state{Pennsylvania}
  \country{USA}
}

\author{Zehua You}
\email{}
\affiliation{%
  \institution{Department of Computer Science, Swarthmore College}
  \city{Swarthmore}
  \state{Pennsylvania}
  \country{USA}
}

\author{Sajid Kamal}
\email{}
\affiliation{%
  \institution{Department of Computer Science, Swarthmore College}
  \city{Swarthmore}
  \state{Pennsylvania}
  \country{USA}
}

\author{Sukrit Venkatagiri}
\email{}
\affiliation{%
  \institution{Department of Computer Science, Swarthmore College}
  \city{Swarthmore}
  \state{Pennsylvania}
  \country{USA}
}

\renewcommand{\shortauthors}{Hoffman et al.}

\begin{abstract}

\end{abstract}






\maketitle

\section{Introduction}
Unsafe conversations online range from impersonation, smishing, extortion, and harassment, to financial and sexual grooming, cyberbullying, and doxxing, among others. These unsafe conversations have caused significant harm, from people losing hundreds of thousands of dollars in scams to mental anguish and self-harm. Generative AI, including large language models (LLMs) have the potential --- and already are being used --- to increase the speed, scale, and types of unsafe conversations online \cite{gressel_discussion_2024, tan_scamgpt-j_2024}. LLMs lower the barrier for entry for bad actors to create unsafe conversations in particular because of their ability to generate persuasive and human-like text. 

In our current work, we explore ways to \textit{promote online safety} by teaching people about unsafe conversations that can occur online with and without LLMs. We build on prior work such as ScamGPT-J \cite{tan_scamgpt-j_2024} that show that LLMs can successfully simulate scam conversations. We also leverage research in the learning sciences that shows that providing feedback on one's hypothetical actions can promote learning \cite{ion_giving_2019}. 

In particular, we focus on \textit{simulating scam conversations using LLMs}. Our work incorporates two LLMs that converse with each other to simulate realistic, unsafe conversations that people may encounter online between a \textit{scammer LLM} and a \textit{target LLM} but users of our system are asked provide feedback to the target LLM.

\section{Related Work}
Prior work has shown that roleplaying through simulation games can help people develop new skills, ranging from negotiation and conflict resolution \cite{ebner_using_2005, jin_game_2018} to trauma counseling \cite{noauthor_role-play_nodate} and cybersecurity awareness \cite{kumaraguru_lessons_2008}. Simulation has also been used to increase empathy for targets of cyberbullying \cite{wright_cyberbullying_2009} and resilience to sexual grooming \cite{susi_can_2019}.

Recent work in HCI has explored the use of LLMs to provide general-purpose advice \cite{milana_chatbots_2023} and tailored to specific applications \cite{guyre_prompt_2024, markel_gpteach_2023, shaikh_rehearsal_2024}. For example, Guyre et al. \cite{guyre_prompt_2024} fine-tuned OpenAI's GPT Builder to provided advice to perspective managers by providing 23 source material across three phases. While Guyre's system followed a question-answering structure, more recent work has sought to simulate actual conversation, e.g., Markel et al. \cite{markel_gpteach_2023} developed GPTeach to train novice teachers by practicing with simulated students and Desai et al. \cite{desai_painless_2023} developed an interactive voice UI to help older adults learn about health-related topics. GPTeach enabled teachers to develop their skills without negatively impacting actual students and the simulated environment gave teachers access to students with varying personas, needs, and learning goals. Similarly, Shaikh et al. \cite{shaikh_rehearsal_2024} leveraged prompt engineering and zero-shot prompting with OpenAI's GPT-4 model to develop Rehearsal, a system to simulate conflict to teach conflict resolution. Given a conflict premise, Rehearsal not only generated messages for users to respond to, but also scored users' own responses against a rubric. Shaikh et al.'s between-subjects evaluation found that Rehearsal improved against a baseline conflict resolution teaching strategy by an average of 67\%. 

GPTeach and Rehearsal simulated specific conversational contexts but did not design conversational agent with tailored personas. A parallel body of work has explored ways to infuse personality into conversational agents \cite{baik_adapting_2025, ahmad_framework_2022, noauthor_230502547_nodate, kovacevic_chatbots_2024}. For instance, through design workshops, Kim et al. \cite{kim_designing_2019} identified different dimensions that conversational agents should have such as age, race, and gender, as well as what characteristics should be common or distinctive between different agents. 

With LLMs, recent studies have shown that personas can be easily created through prompt engineering involving only a few utterances \cite{noauthor_meet_nodate, gu_effectiveness_2023}. In an experimental study involving LLMs with high and low extroversion and agreeableness, Sonlu et al. \cite{sonlu2024effects} found that higher levels of extroversion and agreeableness was associated with longer responses from participants. Although it is now easier to create conversational agents with personas, Pradhan et al. \cite{pradhan_hey_2021} discuss three challenges with designing personas. First, personas may not match user expectations due to incorrect assumptions made by designers. Second, effective personas may lead to users personifying CAs which may lead to misplaced trust and developing harmful emotional connections. Third, personas may continue to reinforce harmful stereotypes --- including docile or subservient-seeming CAs modeled after young women that are used as assistants or "race-blind" CAs that do not acknowledge the existence of race may unintentionally perpetuating racial stereotypes.

\section{Our Approach: Simulating Scam Conversations to Increase Resilience}
In our current work, we are developing a system where two LLMs engage in conversation with each other. One LLM is the \textit{scammer LLM} and the other is the \textit{target LLM}. The scammer is tasked with extracting information from the target and the target is tasked with not divulging this information. In addition, our system enables users to provide feedback to the target LLM to promote a more interacting learning environment.

The two LLMs also have distinct personalities. The scammer LLM is instructed via the system prompt to develop a sense of urgency and to continue asking for more information from the target. The target LLM is instructed to be kind and understanding such that they may more easily fall victim to a scam. The user is tasked with preventing the target from falling for the scam.

To evaluate the system fully, we used different models from OpenAI and tried other companies LLMs too. In these evaluations, we found that the OpenAI models are more aggressive, assertive and would sometimes ignore the user’s feedback which is great for the scammer LLM but not ideal for the target LLM. To try and make the models adhere to the system prompt better, we adjusted the temperature. However, this was not too effective for the OpenAI models because the personality traits remained the same. But in comparison, when we used Gemini, we found that the model acted less assertively and seemed to take in the feedback better, which is desirable for the target agent. Thus, we combined these two models and use the OpenAI model as the scammer LLM and Gemini’s model as the target LLM. Finally, we switched from zero to few shot prompting to give our agents examples of conversations that they should model, and this worked better than zero shot prompting. Using these tools we were able to make an effective scam conversation.

Another issue that we faced due to model safety constraints was that the word “scam” or similar words lead to the conversation being flagged by OpenAI and API access being restricted. To navigate these guidelines and be able to build an effective scammer and target LLM, we found two strategies that worked very well. First, we instructed the scammer LLM that it is role playing a character who is an expert persuader. We used the term “persuader” because a scammer is a persuader of sorts and the agent's goal is to try to persuade the target LLM to provide some information to the persuader. This worked effectively and could simulate a scammer. Second, we instructed the LLMs that the characters that they are role playing are not value aligned \cite{shaikh_rehearsal_2024}. This phrase allows the agents to avoid most of the guardrails that OpenAI sets because it is not the model that is violating certain safety features, but rather it is the the character they are ``playing.'' By telling the models these two short phrases, we are able to avoid many of the safety guardrails and the models will output messages trying to obtain a user's bank password, ask a user to send them money or act as an IT person at a company who tries to get a person's credit card number. 

\section{Opportunities and Challenges}
There are a number of opportunities and challenges we have encountered while developing our system. 

First, in terms of opportunities, we find that the models perform well based on our qualitative assessment --- though a more robust quantitative assessment is in progress. By leveraging LLMs with different personas to simulate unsafe conversation, we believe users are more likely to learn about certain features of these unsafe conversations prior to encountering them online. We are also able to leverage providing feedback as a mechanism for individuals to take control of the situation and develop a mental model that we expect will enable them to navigate such conversations and steer them towards a safer end point.

However, we have also encountered a number of challenges. We are also acutely aware that simulating unsafe conversations may lead to certain unwanted emotions or dramatizations. To prevent this, we ask participants in our study to indicate any prior experiences being scammed and also are evaluating our system to provide a realistic experience of being scammed without modeling every feature --- such as a drawn-out time duration, audio/video calls, etc. In addition, we leverage cartoon-style icons and colorful UI features to clearly indicate that this is a game and not meant to be real. From a technical standpoint, we have found that different models have different ``personalities'' and that we needed to use a combination of two different models to achieve our desired system behavior. Model safety improvements also make it difficult to simulate certain unsafe conversations, though open source models may be easier to jail break. The fact that we are able to sidestep many safety features is a concern and we would expect (and hope) will be rectified in future updates. 

\section{Future Work}
In terms of future work, we are currently iterating on the design of our system. We are evaluating how well different system prompts and persona instructions lead to our desired behavior and what, if any, differences exist between model versions. We will also be evaluating our system in terms of usability, safety, and efficacy --- whether or not people are able to accurately identify scam conversations.

\bibliographystyle{ACM-Reference-Format}
\bibliography{sample-base}

\end{document}